\documentclass[final]{aipproc}

\usepackage{amssymb}
\usepackage{amsmath}
\layoutstyle{6x9}

\begin{document}


\title{The dual Meissner effect in SU(2) Landau gauge}

\author{Tsuneo Suzuki}{address={Institute for Theoretical Physics, Kanazawa University,
Kanazawa 920-1192, Japan\\
and\\
RIKEN, Radiation Laboratory, Wako 351-0158, Japan}}

\author{Katsuya Ishiguro}{address={Institute for Theoretical Physics, Kanazawa University,
Kanazawa 920-1192, Japan}}

\author{Yoshihiro Mori}{address={Institute for Theoretical Physics, Kanazawa University,
Kanazawa 920-1192, Japan}} 

\author{Toru Sekido}{address={Institute for Theoretical Physics, Kanazawa University,
Kanazawa 920-1192, Japan}}

\begin{abstract} 
The dual Meissner effect is observed without monopoles in quenched $SU(2)$ QCD with Landau gauge-fixing. Abelian as well as non-Abelian electric fields are squeezed. Magnetic displacement currents which are time-dependent Abelian magnetic fields play a role of solenoidal currents squeezing Abelian electric fields.  Monopoles are not always necessary to the dual Meissner effect. The squeezing of the electric  flux means the dual London equation and the massiveness of the Abelian electric fields as an asymptotic field.  The mass generation 
of the Abelian electric fields is related to a gluon condensate  $<A^a_{\mu}A^a_{\mu}>\neq 0$ of mass dimension 2.  
\end{abstract}

\maketitle

\section{Introduction}
\begin{figure}[h]
   \begin{minipage}{.5\textwidth}
 \begin{center}
  \includegraphics[width=7cm]{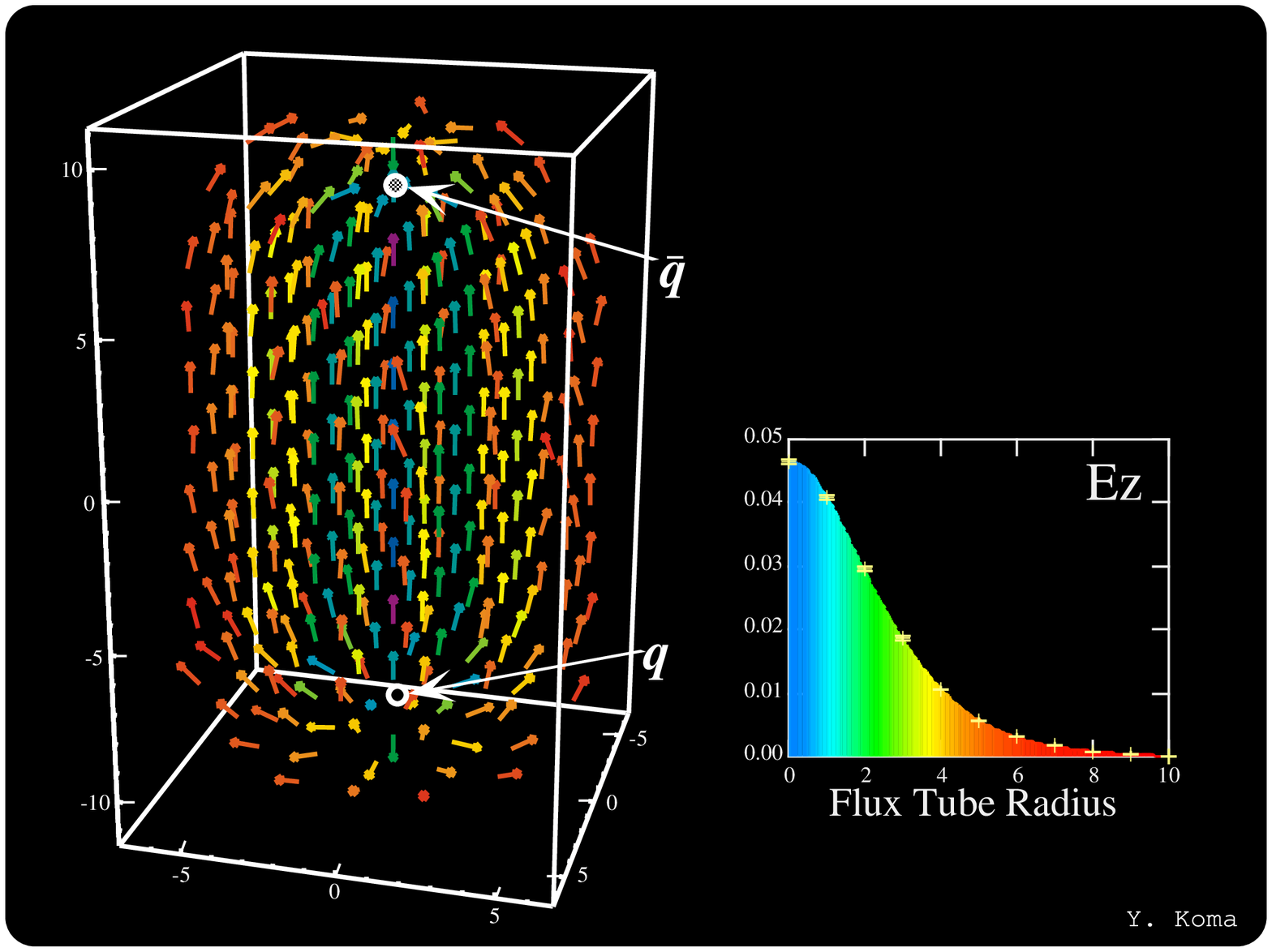}
 \end{center}
   \end{minipage}
   \begin{minipage}{.5\textwidth}
 \begin{center}
  \includegraphics[width=7cm]{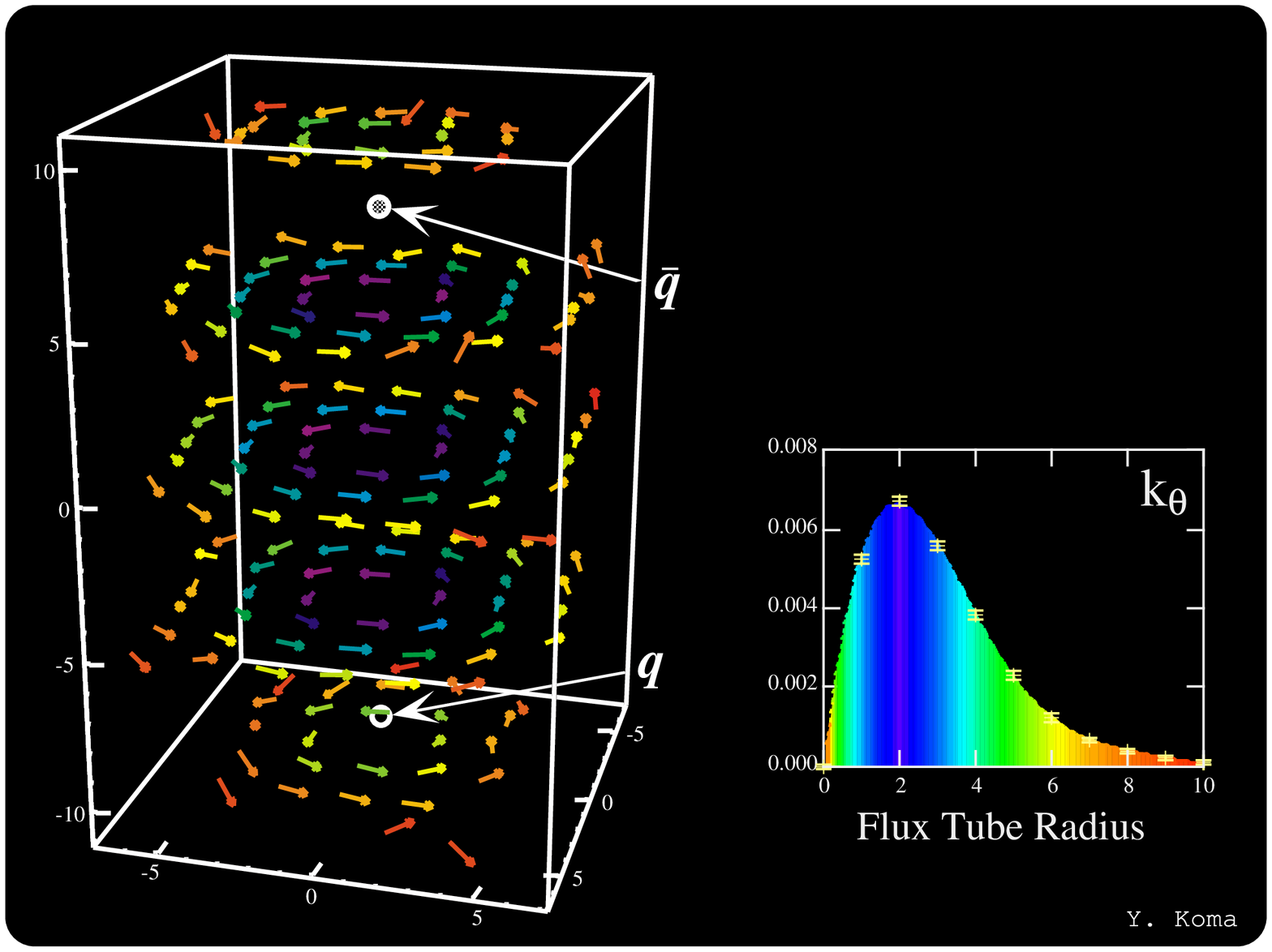}
  \caption{Abelian electric field flux and monopole currents in MA gauge\cite{Koma:2003gq}.}\label{koma1}
 \end{center}
   \end{minipage}
\end{figure}

To understand color confinement mechanism is still an unsolved important problem.
The dual Meissner effect is believed to be the mechanism\cite{tHooft:1975pu, Mandelstam:1974pi}.
However what causes the dual Meissner effect is not clarified. A possible candidate is magnetic monopoles which appear after projecting $SU(3)$ QCD to an Abelian $U(1)^2$ theory by a partial gauge fixing\cite{tHooft:1981ht}. If such monopoles condense, color confinement could be understood due to the dual Meissner effect. Actually  
an Abelian projection adopting a special gauge called Maximally Abelian gauge (MA)\cite{suzuki-83,kronfeld} leads us to interesting results\cite{AbelianDominance,Reviews} which support the idea of monopoles after an Abelian projection. See Fig.\ref{koma1}
in which the dual Meissner effect due to monopole currents as a solenoidal current is seen beautifully\cite{Koma:2003gq}.

Now a natural question arises. What happens in other general Abelian projections or even without any Abelian projection? For example, consider an Abelian projection diagonalizing Polyakov loops. Monopoles exist in the continuum limit at a point where  eigenvalues of Polyakov loops are degenerate\cite{tHooft:1981ht}.
But it is easy to show that such a point runs only in the time-like direction. This means only time-like monopoles which do not contribute to the string tension exist in the Abelian projection\cite{maxim}. Discuss another simple case of Landau gauge. There vacuum configurations are so smooth and it is easy to check numerically that no monopoles coming from singularities exist. Without space-like monopoles or monopoles themselves, monopole condensation could not occur. We have to find another confinement mechanism.   

In this note, we show  that the dual Meissner effect in an Abelian sense works good even when monopoles do not exist, performing  Monte-Carlo simulations of quenched $SU(2)$ QCD with Landau gauge fixing. Instead of monopoles, time-dependent Abelian magnetic fields regarded as magnetic displacement currents are squeezing Abelian electric fields. The dual Meissner effect leads us to the dual London equation and the mass generation of the Abelian electric fields which suggests the existence of a dimension 2 gluon condensate. Our present numerical results are not perfect, since the continuum limit, the infinite-volume limit and the gauge-independence are not studied yet. Moreover our discussions use Abelian components only on the basis of not yet clarified assumption that Abelian components are dominant in the infrared QCD (Abelian dominance\cite{AbelianDominance,Reviews,greensite-96,Cea:1995zt}). Nevertheless authors think the present results are very interesting to general readers, since they show for the first time the Abelian dual Meissner effect is working in lattice non-Abelian QCD without resorting to monopoles coming from a singular gauge transformation\cite{faber}. The gauge adopted here is the simplest one only to get smooth configurations.   The present results hence suggest the Abelian dual Meissner effect is the real universal mechanism of color confinement which has been sought for many years. Moreover
the relation of the Abelian dual Meissner effect with the dimension 2 gluon condensate 
sheds new light on the importance of the gluon condensate
\cite{zakharov,kondo,dudal,arriola,slavnov}. 
Detailed studies, a direct proof of Abelian dominance and extensions to $SU(3)$ and full QCD in zero and finite-temperature cases are in progress.  

\begin{figure}
\includegraphics[height=6cm, width=8.5cm]{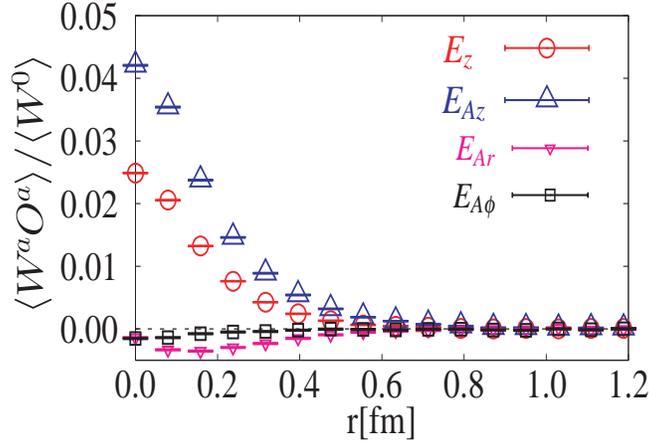}
\caption{\label{fig_1} Abelian $\vec{E}_A$ and non-Abelian $\vec{E}$ electric field  profiles in Landau gauge. $W(R\times T= 6\times 6)$ is used.}
\end{figure}

\section{Simulation details}
As a lattice quenched $SU(2)$ QCD,   
we use an improved gluonic action found by Iwasaki\cite{iwasaki}.
\begin{eqnarray}
S = \beta \left\{c_0 \sum Tr (plaquette) + c_1 \sum Tr (rectangular) \right\},\label{iwasaki}
\end{eqnarray}
with which better scaling behaviors of physical quantities are expected.
The mixing parameters are fixed as 
$c_0 + 8c_1 = 1$ and  $c_1=-0.331$.

To measure correlations of gauge-variant electric and magnetic fields directly, we 
adopt a simplest gauge  called  Landau gauge   which maximizes
$\sum_{s,\mu} Tr [ U_{\mu}(s) + U_{\mu}^{\dagger}(s)]$.
After the gauge fixing, we try to measure electric and magnetic flux distributions by evaluating correlations of Wilson loops and field strengths. For comparison, we also 
use MA gauge where 
$\sum_{s,\mu} Tr [ U_{\mu}(s)\sigma_3U_{\mu}^{\dagger}(s)\sigma_3]$
is maximized.
To get a good signal to noise ratio,  the APE smearing technique\cite{APE}
is used when evaluating Wilson loops $W(R, T)=W^0+iW^a\sigma^a$:
\begin{eqnarray}
U_i(s) \longrightarrow N \left\{ U_i(s) + \alpha \sum_{j \neq i} U_j(s)U_i(s+\hat{j})U_j^\dagger
(n+\hat{i})\right\},
\end{eqnarray}
where $N$ is normalization factor and $\alpha$ is a free parameter. We have used  $\alpha = 0.2$  and $N=80$.  

Measurements of the string tension make us  fix the scale when we use $\sqrt{\sigma}=440${MeV}.
We adopt a coupling constant $\beta=1.2$ in which the lattice distance $a(\beta=1.2)$ is $0.07921(22)$[fm]. The scale is chosen only because we  compare our results with those studied extensively in MA gauge\cite{bali-96,Koma:2003gq}.
The lattice size is $32^4$ and after 5000 thermalization, we have prepared  
5000 thermalized configurations per each 100 sweeps for measurements.

Non-Abelian electric and magnetic fields  are defined from $1\times 1$ plaquette 
$U_{\mu\nu}(s)=U^0_{\mu\nu}+iU^a_{\mu\nu}\sigma^a$ as done in Ref.\cite{bali-94}: 
\begin{eqnarray}
E^a_k(s)&\equiv& \frac{1}{2}(U^a_{4k}(s-\hat{k})+U^a_{4k}(s))\nonumber\\
B^a_k(s)&\equiv& \frac{1}{8}\epsilon_{klm}(U_{lm}^a(s-\hat{l}-\hat{m})\nonumber \\
&&+U_{lm}^a(s-\hat{l})+U_{lm}^a(s-\hat{m})+U_{lm}^a(s))\nonumber
\end{eqnarray}
We also define Abelian electric ($E_{Ai}^a$) and magnetic fields ($B_{Ai}^a$) similarly using Abelian plaquettes $\theta_{\mu\nu}^a(s)$ defined through  link variables $\theta_{\mu}^a(s)$: 
\begin{eqnarray}
\theta_{\mu\nu}^a(s)\equiv\theta_{\mu}^a(s)+\theta^a_{\nu}(s+\hat{\mu})-\theta_{\mu}^a(
s+\hat{\nu})-\theta^a_{\nu}(s)
\end{eqnarray}
where $\theta_{\mu}^a(s)$ is given by 
$U_{\mu}(s)=\exp(i\theta_{\mu}^a(s)\sigma^a)$.
In MA gauge, the Abelian link variables $\theta_{\mu}^{MA}(s)$ are defined by a phase of the diagonal part of the non-Abelian link field.
\begin{eqnarray*}
U_{\mu}^0(s)&=&\sqrt{1-|c_{\mu}(s)|^2}\cos\theta^{MA}_{\mu}(s),\\
U_{\mu}^3(s)&=&\sqrt{1-|c_{\mu}(s)|^2}\sin\theta^{MA}_{\mu}(s).
\end{eqnarray*}
Since the off-diagonal part $|c_{\mu}(s)|$ is small\cite{AbelianDominance},  $\theta_{\mu}^{MA}(s)\sim \theta^3_{\mu}(s)$ in MA gauge. 
As a source corresponding to a static quark and antiquark pair,
we use here only non-Abelian Wilson loops.

\section{Results}

First we show in Fig.\ref{fig_1} Abelian and non-Abelian electric   flux profiles around a pair of static quark and antiquark in Landau gauge. The profiles are mainly studied on the perpendicular plane at the midpoint between the quark pair. Note that electric fields perpendicular to the $Q\bar{Q}$ axis are found to be negligible. It is very interesting to see from Fig.\ref{fig_1} that Abelian electric field $E_{Az}$ simply defined here is squeezed also. Moreover the squeezing of $E_{Az}$  is  stronger than that of non-Abelian one $E_z$.  To know how squeezing of the Abelian flux occurs seems hence essential.

\begin{figure}[h]
   \begin{minipage}{.5\textwidth}
 \begin{center}
  \includegraphics[width=7cm]{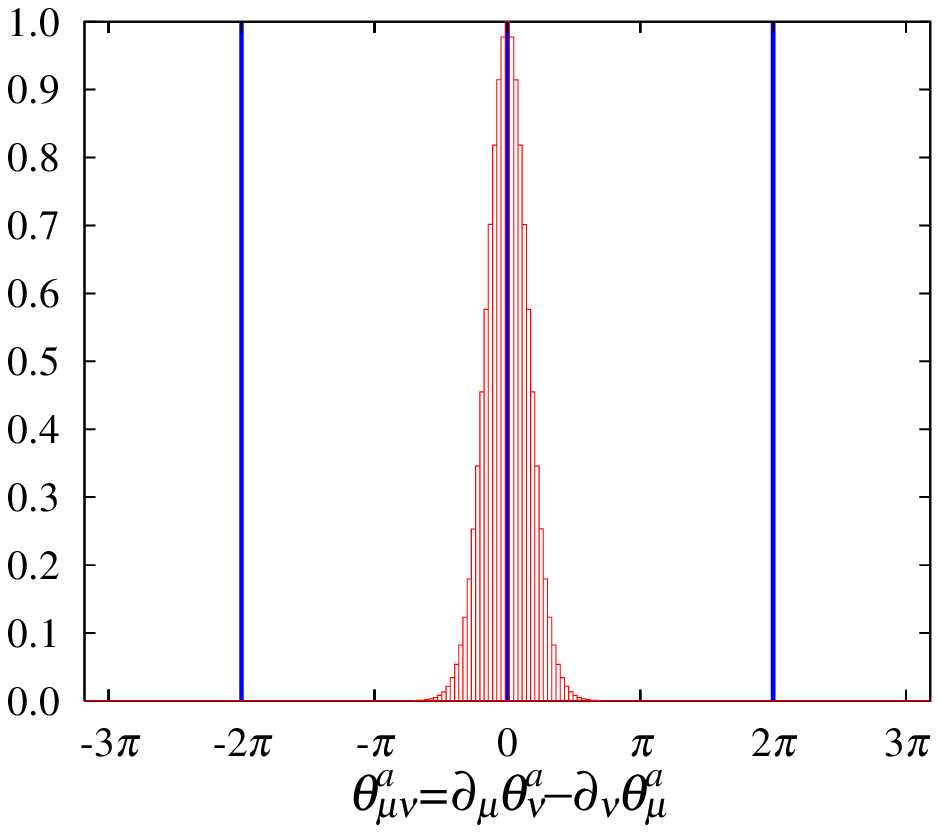}
 \end{center}
   \end{minipage}
   \begin{minipage}{.5\textwidth}
 \begin{center}
  \includegraphics[width=7cm]{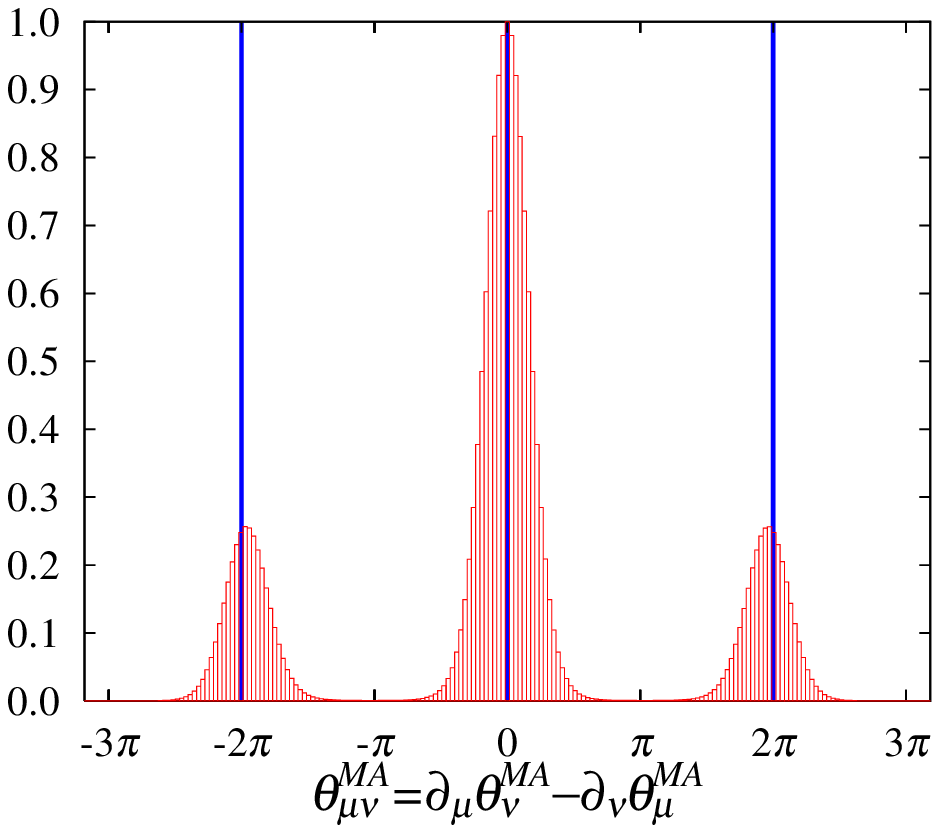}
  \caption{Histograms of $\theta_{\mu\nu}$ in Landau (Left) and MA (Right) gauges}\label{fmunu}
 \end{center}
   \end{minipage}
\end{figure}

Let us discuss from now on flux distributions of Abelian  fields alone. 
It is checked numerically that  there are no DeGrand-Toussaint monopoles\cite{DeGrand:1980eq}. See Fig.\ref{fmunu} in which histograms of Abelian field strength
$\theta_{\mu\nu}$ are plotted in Landau gauge (Left) and MA gauge (Right).

 Hence the Abelian fields satisfy kinematically the simple Abelian Bianchi identity:
\begin{eqnarray}
\vec{\nabla}\times\vec{E}_{A}^a=\partial_{4}\vec{B}_{A}^a, \hspace{1cm}
\vec{\nabla}\cdot\vec{B}_{A}^a=0.\label{BI1}
\end{eqnarray}
In the case of MA gauge, there are additional monopole current $(\vec{k}, k_4)$ contributions:
\begin{eqnarray}
\vec{\nabla}\times\vec{E}^{MA}=\partial_{4}\vec{B}^{MA}+\vec{k}, \hspace{1cm}
\vec{\nabla}\cdot\vec{B}^{MA}=k_4.\label{BI2}
\end{eqnarray}
Here $\vec{E}^{MA}$ and $\vec{B}^{MA}$ are defined in terms of plaquette variables 
$\theta_{\mu\nu}^{MA}(s)\ \  (\textrm{mod}\ \ 2\pi)$ which are constructed by  $\theta^{MA}_{\mu}(s)$. 
 
The Coulombic electric field coming from the static source is written in the lowest perturbation theory in terms of the gradient of a scalar potential. Hence it does not contribute to the curl of the Abelian electric field nor to the Abelian magnetic field in 
the above Abelian Bianchi identity  Eq.(\ref{BI1}). 
The dual Meissner effect  says that the squeezing of the electric flux occurs due to cancellation of the Coulombic electric fields and those from  solenoidal magnetic currents. In the case of  MA gauge, magnetic monopole currents $\vec{k}$ play the role of the solenoidal current\cite{Singh:1993jj,bali-96,Koma:2003gq}. 

\begin{figure}
\includegraphics[height=6cm]{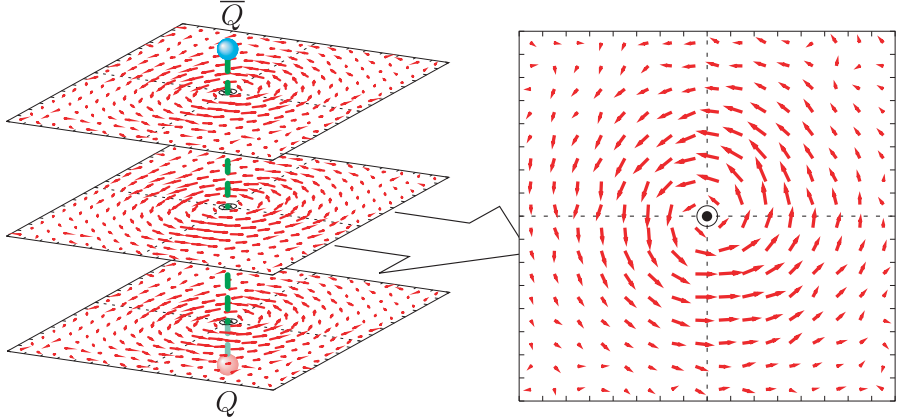}
\caption{\label{fig_2}Magnetic displacement currents in Landau gauge as a solenoidal current.}
\end{figure}

 \begin{figure}[h]
   \begin{minipage}{.5\textwidth}
 \begin{center}
  \includegraphics[width=7cm]{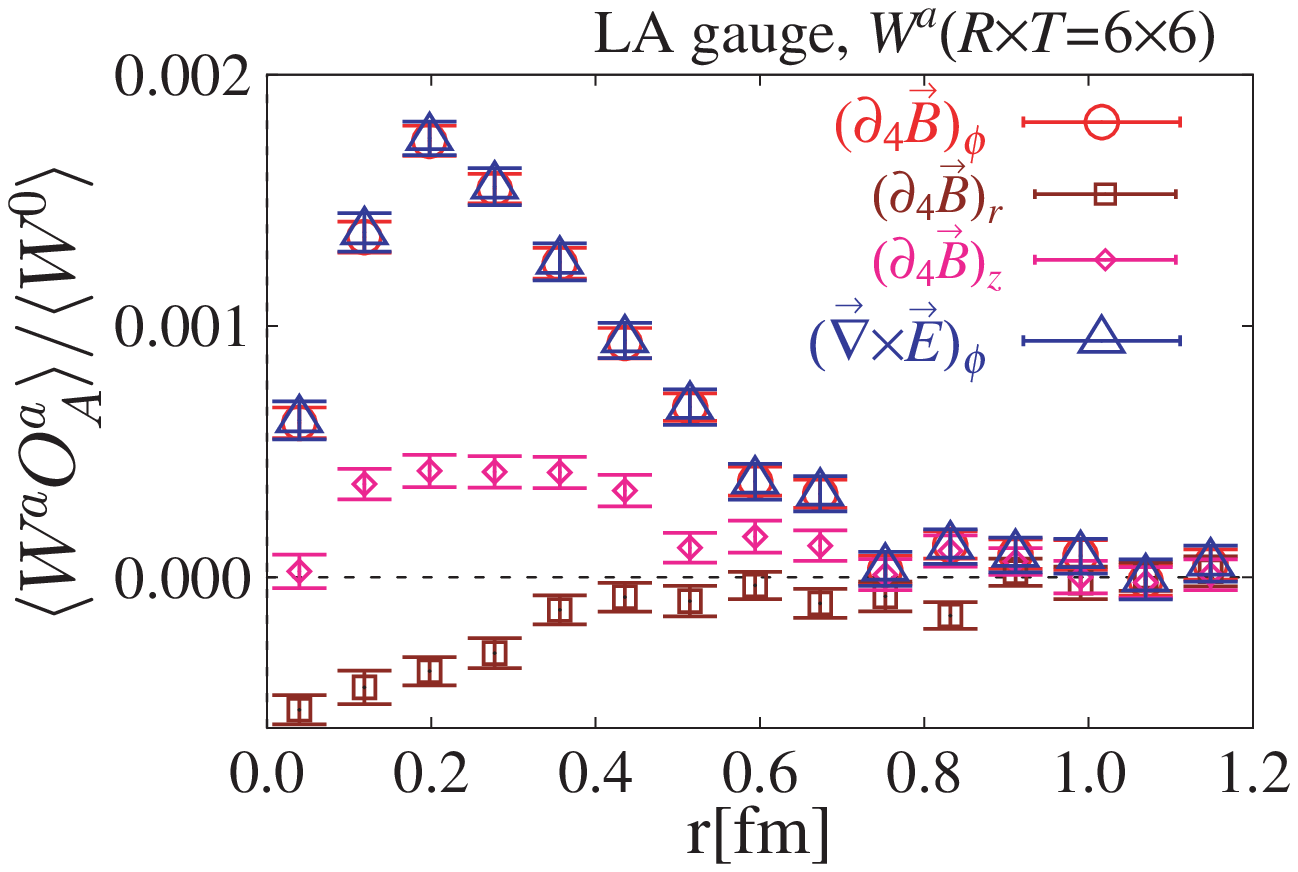}
 \end{center}
   \end{minipage}
   \begin{minipage}{.5\textwidth}
 \begin{center}
  \includegraphics[width=7cm]{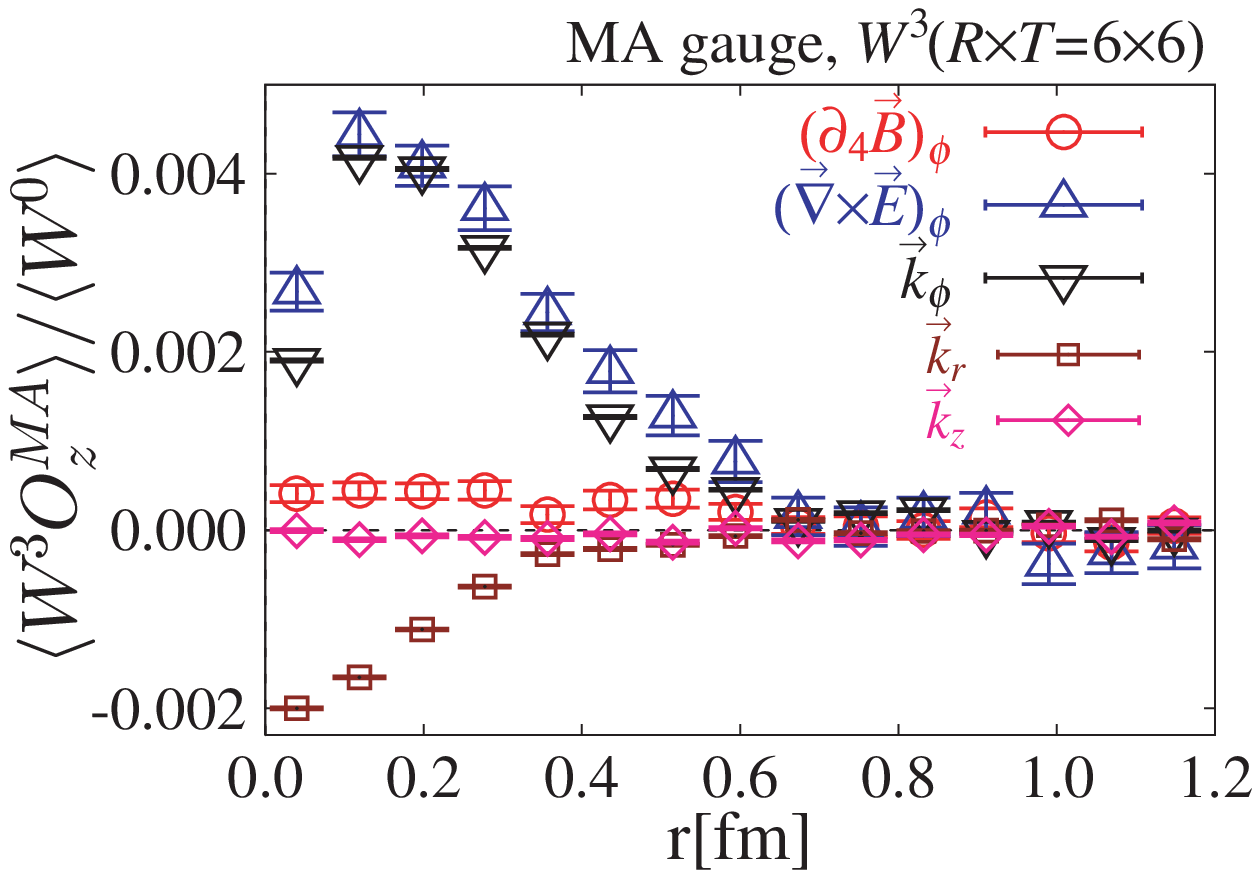}
  \caption{Curl of Abelian electric fields and magnetic displacement currents around a static quark pair in Landau (Left) and in MA (Right) gauges. Monopole currents are also plotted in MA gauge.}\label{curl}
 \end{center}
   \end{minipage}
\end{figure}

Now what happens in a smooth gauge like the Landau gauge where monopoles do not exist?
 From Eq.(\ref{BI1}), only $\partial_4\vec{B}_{A}$ regarded as a magnetic displacement  current could play the role of the solenoidal  current. It is very interesting to see  Fig.\ref{fig_2}  in which this happens actually in Landau gauge. Note that the solenoidal current has  a direction squeezing the Coulombic electric field. Let us see also the detailed distributions shown in Fig.\ref{curl}. The other components of the magnetic displacement current $\partial_4B_{Ar}$ and $\partial_4B_{Az}$ are not vanishing but they are much suppressed consistently with 
 Fig.\ref{fig_1}.  In comparison, we show the case of MA gauge also in Fig.\ref{curl}. Here  $\partial_4B_{A\phi}$ is found to be negligible numerically as already expected  from the works\cite{Singh:1993jj,Cea:1995zt,bali-96}. Instead monopole currents are circulating\cite{Singh:1993jj,bali-96,Koma:2003gq}. In this case, $k_r$ is non-vanishing, although it is also suppressed
 in comparison with $k_{\phi}$. $k_z$ is almost zero.
 The authors think that non-vanishing of the radial and $z$ components of $\partial_4\vec{B}$ in Landau gauge and $\vec{k}$ in MA gauge is  due to lattice artifacts and also due to the smallness of the Wilson loop size adopted here.
It is interesting that the shapes of $\partial_4B_{A\phi}$ in Landau gauge and $k_{\phi}$ in MA gauge look similar, although the strengths are different. They have a peak at almost the same 
distance around $0.2$[fm] and almost vanish around $0.7$[fm].  

\section{Abelian dominance test}
The reader may wonder if the above consideration of the Abelian fields alone is enough. Some people believe that the non-perturbative confinement problem in infrared QCD could be understood in terms of Abelian quantities\cite{AbelianDominance,Reviews,kondo-04}. 

\begin{figure}
\includegraphics[height=6cm]{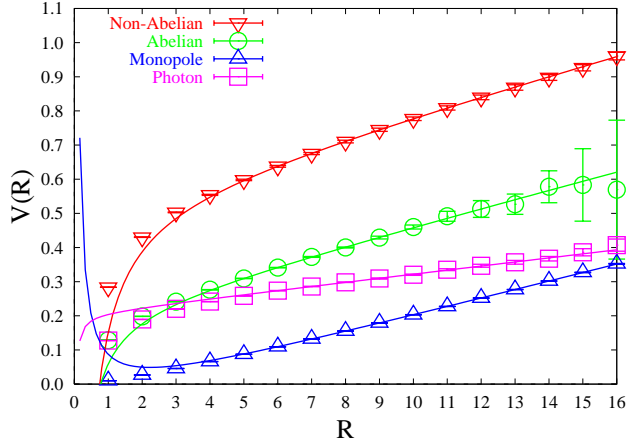}
\caption{\label{potential}Abelian, monopole and photon static potentials.}
\end{figure}

First we evaluate the Abelian and monopole contributions to the string tension in MA gauge using the Iwasaki improved action (\ref{iwasaki}). It is plotted in Fig.\ref{potential}. Our results are as follows:
\begin{eqnarray*}
\frac{\sigma_{ab}}{\sigma}&=&0.94\pm0.05,\ \ \ 
\frac{\sigma_{mo}}{\sigma_{ab}}=0.98\pm0.02,\ \ \ 
\frac{\sigma_{mo}}{\sigma}=0.92\pm0.01.
\end{eqnarray*}
This is compared with the result obtained using Wilson gauge action in Ref.\cite{bali-MA}:
\begin{eqnarray*}
\frac{\sigma_{mo}}{\sigma}
&=&0.87\pm0.02
\end{eqnarray*}
Hence substantial improvement is obtained with the use of Iwasaki impoved action.
This suggests the abelian monopole part could reproduce the full string tension
in the continuum limit.

\begin{figure}
\includegraphics[height=6cm]{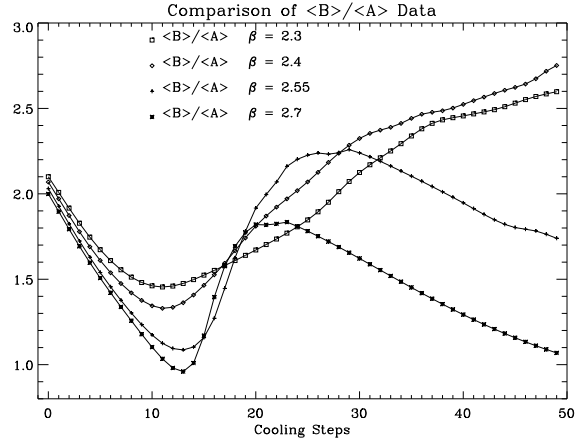}
\caption{\label{greensite}Non-abelianicity $Q$ vs. Cooling Step, for several
values of $\beta$. This figure is from Ref.\cite{greensite-96}.}
\end{figure}

Abelian dominance in infrared QCD is also checked in some works with the use of 
controlled cooling. Giedt and Greensite
\cite{greensite-96} have measured the ratio $B/A$ where 
\begin{eqnarray*}
B &=& -{1 \over V} \sum_x {1 \over n_p (n_p - 1)} \sum_{i>j} \sum_{m>n} \mbox{Tr} \{ [F_{ij}(x),F_{mn}(x)]^2 \}\\ 
A &=& {1 \over n_p V} \sum_x \sum_{i>j} \mbox{Tr}\{ F_{ij}^4 \}
\end{eqnarray*}
and have shown that it decreases as the cooling steps as far as the string tension is kept non-vanishing. See Fig.\ref{greensite} taken from Ref.\cite{greensite-96}.
Cooling is expected to reduce the big Coulombic interaction in the non-Abelian case while keeping infrared confinement property.

\begin{figure}
\includegraphics[height=6cm]{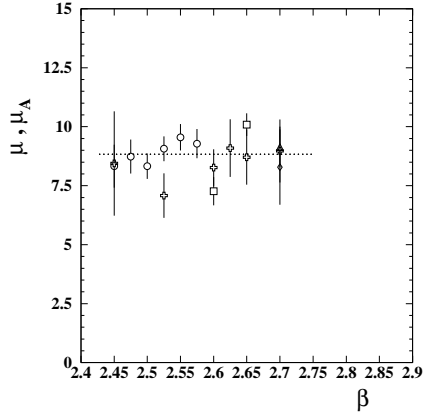}
\caption{\label{cea}$\mu$ and $\mu_A$ (in units of $\Lambda_{\overline{MS}}$)
versus $\beta$  for square Wilson loops. Circles, squares, and
triangle refer to $L=16$, $20$, $24$ respectively. Crosses and diamond
refer to the Abelian projected correlator $\rho^A_W$ with $L=16$, $20$
respectively.This figure is from Ref. \cite{Cea:1995zt}.}
\end{figure}

Cea and Cosmai\cite{Cea:1995zt} have measured connected correlation operators of Wilson loops and $1\times 1$ plaquette using the controlled cooling and have obtained the penetration length in non-Abelian case. It is almost equal to the penetration length determined 
by Abelian Wilson loop and Abelian electric fields in MA gauge. See Fig.\ref{cea} taken from Ref.\cite{Cea:1995zt}.

We also try to check Abelian dominance in Landau gauge using a controlled cooling \cite{cooling} under which  the string tension remains non-vanishing. 
For reader convenience let us, briefly, illustrate our cooling
procedure. The lattice  gauge configurations are cooled by replacing the
matrix $U_\mu(s)$ associated to each link $l\equiv(s,\hat{\mu})$ with
a new matrix $U_\mu^\prime(s)$ in such a way that the local
contribution to the lattice action
\begin{equation}
\label{DeltaS}
S(s) = 1 - \frac{1}{2} \text{tr} \left\{ U_\mu(s) k(s) F(s)
\right\}
\end{equation}
is minimized. $\widetilde{F}(s) = k(s) F(s)$ is the sum over the
``U-staples'' involving the link $l$ and
$k(s)=\sqrt{\text{det}\left(\widetilde{F}(s)\right)}$, so that
$F(s)\in$~SU(2). In a ``controlled'' or ``smooth'' cooling step we
have
\begin{equation}
\label{controlled_cooling}
U_\mu(s) \rightarrow  U^\prime_\mu(s) = V(s) U_\mu(s) \;,
\end{equation}
where $V(s)$ is the SU(2) matrix which maximizes 
\begin{equation}
\label{Vmax}
\text{tr} \left\{ V(s) U_\mu(s) F(s) \right\}
\end{equation}
subjected to the following constraint on the SU(2) distance between
$U_\mu(s)$ and $U^\prime_\mu(s)$:
\begin{equation}
\label{constraint}
\frac{1}{4} \text{tr} \left[ \left( U_\mu^\dagger(s) -
U^{\prime\dagger}_\mu(s) \right)  \left( U_\mu(s) - U^\prime_\mu(s)
\right) \right] \; \le \delta^2 \,.
\end{equation}
We adopt $\delta=0.3$. A complete cooling sweep consists in the
replacement Eq.~\eqref{controlled_cooling} at each lattice site. 

\begin{figure}
\includegraphics[height=6cm, width=8.5cm]{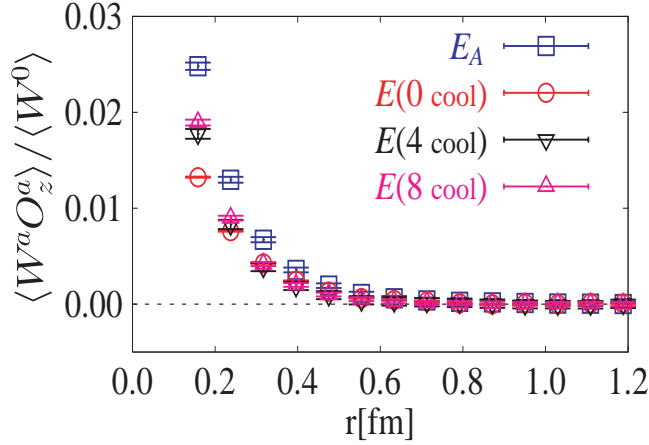}
\caption{\label{fig_5}Profiles of non-Abelian electric field $E_z^a$ after cooling. $E_{Az}^a$ stands for the Abelian electric field. $W(R\times T= 6\times 6)$ is used.}
\end{figure}

We find the profile of the non-Abelian electric field $E_{z}^a$ tends to that of the Abelian one of $E_{Az}^a$ in the long-range region as shown in Fig.\ref{fig_5}. This is consistent
with the above result\cite{Cea:1995zt} in a different approach. 

\section{Dimension 2 gluon condensate}
 
Now we have shown that the magnetic  displacement currents are important in the dual Meissner effect when there are no monopoles. Then how can we understand the origin of the dual Meissner effect without monopoles?
The Abelian dual Meissner effect indicates the massiveness of the Abelian electric field as an asymptotic field:
\begin{eqnarray}
(\partial_{\rho}^2-m^2)\vec{E}_A\sim 0.\label{eqm-abel}
\end{eqnarray}
This leads us to  a dual London equation which is a key  to the dual Meissner effect. Let us evaluate the curl of the magnetic  displacement current. Using Eq.(\ref{BI1}), we get
\begin{eqnarray}
\vec{\nabla}\times\partial_4\vec{B}_A&=&\vec{\nabla}(\vec{\nabla}\cdot\vec{E}_A)-\vec{\nabla}^2\vec{E}_A.\nonumber 
\end{eqnarray}
From Eq.(\ref{eqm-abel}), we get the dual London equation:  
\begin{eqnarray}
\vec{\nabla}\times\partial_4\vec{B}_A\sim (\partial_4^2-m^2)\vec{E}_A.\label{London2}
\end{eqnarray}

Let us remember a simple mean-field approach developed by Fukuda\cite{fukuda-82}.
Neglecting gauge-fixing and Fadeev-Popov terms, we have 
equations of motion
$D_{\mu}^{ab}F_{\mu\nu}^b= 0$ 
and the (non-Abelian) Bianchi identity
$D_{\mu}^{ab}{}^*F_{\mu\nu}^b=0$.
Applying $D$ operator to the Bianchi identity and using the Jacobi identity and the equations of motion,  we get 
\begin{eqnarray}
(D_{\rho}^2)^{ab}F_{\mu\nu}^b&=&2g\epsilon^{abc}F_{\mu\alpha}^bF_{\nu\alpha}^c.
\label{LO}
\end{eqnarray}
Notice
$(D_{\rho}^2)^{ab} = \partial_{\rho}^2\delta^{ab}+g\epsilon^{acb}(\partial_{\rho}A_{\rho}^c)+g^2(A_{\rho}^aA_{\rho}^b-\delta^{ab}(A_{\rho}^c)^2)$. 
Hence if $<A^a_{\mu}A^b_{\nu}>=\delta^{ab}\delta_{\mu\nu}v^2\neq 0$, we see 
asymptotically that the electric fields become massive $(\partial_{\rho}^2-m^2)E^a_k\sim 0$
with $m^2=8g^2v^2$\cite{mass}. Now the Abelian electric field is also massive asymptotically
$(\partial_{\rho}^2-m^2)E^a_{Ak}\sim 0$.
Hence the dual London equation (\ref{London2}) is  obtained.

The importance of the dimension 2 gluon condensate has been stressed by Zakharov and his collaborators\cite{zakharov} and Refs.\cite{kondo,dudal}. Recent discussions on the value of the gluon condensate are seen in Ref.\cite{arriola}. Some of them discuss the mass generation of the gluon propagator. But as pointed out by Fukuda\cite{fukuda-82}, the non-vanishing dimension 2 gluon condensate leads us to a conclusion that field strengths $F_{\mu\nu}^a(s)$ instead of gluon fields $A_{\mu}^a(s)$ become  good canonical variables having a finite mass, whereas  gluon propagators have a $(p^2)^{-2}$ behavior showing confinement. 

Although the operator of the gluon condensate is gauge-variant, it is proved recently 
the expectation value is gauge invariant\cite{slavnov}. Hence the gluon condensate has a physical importance, if the proof is correct. 

\begin{theacknowledgments}
The numerical simulations of this work were done using RSCC computer clusters in 
RIKEN. The authors would like to thank RIKEN for their support of computer facilities. 
T.S. is supported by JSPS Grant-in-Aid for Scientific Research on Priority Areas 13135210 and (B) 15340073. T.S. is thankful to Valentine Zakharov, Mikhail Polikarpov and Manfred Faber for useful discussions.
\end{theacknowledgments}


\begin{thebibliography}{99}
\bibitem{mass}
A mass counter term is needed rigorously.
\end{thebibliography}
\end{document}